\documentclass[aps,prd,twocolumn,nofootinbib,longbibliography]{revtex4-2}
\pdfoutput=1
\usepackage[english]{babel}
\usepackage[autostyle, english = american]{csquotes}
\usepackage[CJKbookmarks, pdftex, bookmarksnumbered, bookmarksopen, colorlinks, citecolor=cyan, linkcolor=cyan,urlcolor=cyan]{hyperref}
\usepackage{amsmath,amssymb, dsfont}
\usepackage{graphicx}
\usepackage{enumitem}
\setlist[enumerate]{topsep=0pt,parsep=-1mm,leftmargin=5mm,}
\MakeOuterQuote{"}
\def\be{\begin{equation}}
\def\ee{\end{equation}}

\newcommand{\ket}[1]{{| #1 \rangle}}

\newcommand{\ketbra}[2]{{| #1 \rangle\!\langle #2 |}}

\renewcommand{\S}{{\mathcal S}\xspace}
\newcommand{\F}{{\mathcal F}\xspace}
\newcommand{\W}{{\mathcal W}\xspace}
\renewcommand{\O}{{\mathcal O}\xspace}
\renewcommand{\H}{{\mathcal H}\xspace}
\newcommand{\A}{{\mathcal A}\xspace}
\newcommand{\B}{{\mathcal B}\xspace}

\renewcommand{\H}{{\mathcal H}\xspace}

\newcommand{\SF}{{\S{-}\F}\xspace}
\newcommand{\SFW}{{\S{-}\F{-}\W}\xspace}

\DeclareMathOperator{\tr}{tr}

\usepackage{xcolor}
\usepackage{xspace}
\usepackage{ifthen}
\newcommand{\showcomments}{true}

\newcommand{\andrea}[1]%
{\ifthenelse{\equal{\showcomments}{true}}%
{{\color{orange}{\small \textbf{A:} #1}}}{\xspace}}%

\newcommand{\carlo}[1]%
{\ifthenelse{\equal{\showcomments}{true}}%
{{\color{blue}{\small \textbf{C:} #1}}}{\xspace}}%

\begin{document}

\title{Relative Information, Relative Facts}

\date{\today}

\author{Andrea {Di Biagio}}
\affiliation{Institute for Quantum Optics and Quantum Information (IQOQI) Vienna, Austrian Academy of Sciences, Boltzmanngasse 3, A-1090 Vienna, Austria}
\affiliation{Basic Research Community for Physics e.V., Mariannenstraße 89, Leipzig, Germany}

\author{Carlo Rovelli}
\affiliation{Aix-Marseille University, Universit\'e de Toulon, CPT-CNRS, F-13288 Marseille, France}
\affiliation{The Rotman Institute of Philosophy, 1151 Richmond St.~N London  N6A5B7, Canada}

\begin{abstract}
\noindent We offer a fresh perspective on the relational interpretation of quantum mechanics as a way of thinking about the world described by quantum theory based on quantifiable notions of information. This allows us to provide a definition of a relative fact, with no addition to orthodox quantum theory and no fundamentally special role for observers. By associating perspectives with commutative observables rather than entire quantum systems, several previous problems with the interpretation are dissolved. As a side result, we show how a quantum measurement, properly described, is a continuous process.
\end{abstract}

\maketitle

\section{A way of thinking}
\label{sec:thinking}

\noindent The task of finding an interpretation of quantum theory is sometimes presented as looking for an \textit{ontology} to complement the formulas and empirical predictions of quantum mechanics, an account of how the world \textit{really is}, beyond our observations. But there is an alternative way of understanding  the task of an interpretation: that of offering a fertile \textit{way of thinking} about nature, to articulate a coherent conceptual structure for our current best understanding of it. Taking an ontology seriously is, afterall, one way of thinking about the world, not necessarily the best one.

For this, we take a fresh look at the relational interpretation of quantum mechanics (RQM)~\cite{rovelli2021relational,sep-qm-relational}. RQM holds that the key to understanding quantum phenomena is to think in terms of the information that different parts of nature have about each other~\cite{rovelli1996relational}. While the idea that quantum theory is about information has a long history~\cite{wheeler1986how,zeilinger1999foundational,fuchs2002quantum,hardy2001quantum,brukner2003information}, in RQM we want to think about information \textit{any} system may have~\cite{rovelli1996relational}, not only macroscopic observers~\cite{fuchs2007quantum} or decision-making agents or individuals~\cite{fuchs2019qbism,healey2010quantum}. This way of thinking about physics avoids both a simple realism (the world is in a definite state) and instrumentalism (we are only allowed to talk about what we, as rational beings, measure).

The main observation on which this paper is based is that probability theory allows for precise quantitative definitions of the information that we, or any other physical system, might have about a certain physical variable. A precise definition of "fact" can be given in these terms, as occurrences about which information is maximal. Building on these, we propose a definition of a ``relative fact''~\cite{brukner2020facts,dibiagio2021stable}. Applying these notions to the standard quantum formalism leads to a way of thinking about quantum theory that allows one to talk of facts in nature and of the information about parts of nature that that we, as well as any other part of nature, may have.  

\medskip

This perspective allows us to take a fully naturalistic standpoint.  By this we mean that we assume that we, sense-making, reasoning, thinking, feeling, human animals are physical entities like any other physical entity. We are distinguished only in the way that our physical makeup is intricately arranged to behave in all these special ways. From this point of view, whatever an observer does, thinks, and knows---including our own theorising---is itself a physical process, one that can, at least in principle, be accounted for by our physical theories.

Specifically, the knowledge that we have about nature and its facts does not live in an abstract realm outside nature itself, but it is embodied in our very physical configuration, which is in principle concretely accessible to other systems.  The information that \textit{we} have about the world is understood as a special case of the information that parts of nature have about one another, where "information" is quantifiable in its elementary physical sense based on correlations.

\section{Relative information}
\label{sec:info}

\noindent We start with some simple definitions, based on Shannon's theory of information~\cite{shannon1948mathematical}, for the concepts of information, mutual information, and facts, grounded on probability theory. This section and the next one do not specifically refer to quantum theory and are independent of any particular interpretation of probability. 

Say the outcome of a measurement of a variable $A$ will yield the value $a$ with probability $p(a)$.  The more sharply peaked $p(a)$ is, the larger we say the \textit{information} about $A$ is.  Following Everett~\cite{long-thesis,everett2012long}, we quantify the amount of information about this variable as
\begin{equation}
  I_A:= I^{\max}_A -H_A,
  \label{const}
  \end{equation}
where $H_A$  is the Shannon entropy of the probability distribution
\begin{equation}
 H_A :=-\sum_a p(a)\log p(a),
\end{equation}
and $I_A^{\max}=\log N_A$ is the logarithm of the number of values $A$ can take.%
\footnote{
Everett~\cite{long-thesis} actually defined $I_A=-H_A$, so that $I^{\max}_A=0$. His definition has the advantage of not depending on the number of possible values of the variable and thus can be naturally extended to continuous variables, but then  $I_A\leq0$, which is a little awkward. Our choice of $I^{\max}_A=\log N_A$ makes $I_A\geq0$. Note that the value of $I_A^{\max}$ does not matter much, as we are interested mostly in differences in information. Additionally, all our definitions are independent of this choice.} 
The Shannon entropy ${H_A=I^{\max}_A-I_A}$ measures the \textit{lack} of information about the variable $A$, it vanishes when $p(a)$ is peaked on one value and it grows as $p(a)$ becomes wider. The information $I_A$ is instead minimal when nothing is known about $A$ and grows as the distribution gets more peaked, reaching its maximum value when $p(a)$ is peaked on a single value of $A$.

Now consider a second variable, $B$, possibly belonging to a different system, and a joint probability distribution $p(a,b)$ over the values of $A$ and $B$. We say that the two variables are \textit{correlated} if $p(a,b)$ is different from the product of its two marginals ${p(a)=\sum_b p(a,b)}$ and ${p(b)=\sum_a p(a,b)}$. We can quantify the correlations by the well-known \textit{mutual information}:
\begin{equation}
I_{A:B}:= H_A+H_B-H_{AB}=I_{AB}-I_A-I_B.
\end{equation}
If $A$ and $B$ are correlated, we can obtain information about $A$ by obtaining information about $B$. This can be made quantitative, as follows.

Say we obtain the value $b$ of $B$, then the probability of the value of $A$ gets updated from the marginal $p(a)$ to the conditional
\begin{equation}
  p(a|b) = \frac{p(a,b)}{p(b)}\,.
\end{equation}
The information about $A$ \textit{conditional} on $B$ taking the value $b$ is, then, 
\begin{equation}
  I_{A|b} := I^{\max}_A+ \sum_{a}p(a|b)\log p(a|b).
\end{equation}
The information about $A$ does not decrease upon learning about $B$, in the sense that $I_{A|b}\geq I_A$ for all $b$, but the increase in information depends on the value $b$. Let us denote by $I_{A|B}$ the expectation value of the conditional information,
\begin{equation}
I_{A|B}:=\langle I_{A|b}\rangle  = \sum_b p(b)I_{A|b}.
\end{equation}
Then one can show the following key relation
 \begin{equation}\label{key}
    I_{A|B}= I_A+I_{A:B},
\end{equation}
that is, the mutual information is {\em the expectation value of the increase in information about $A$ after looking at $B$.}  

In the rest of the work, we will call $I_{A|B}$ the information about $A$ \textit{relative to} $B$. In terms of the joint probability distribution $p(a,b)$, the relative information can be written as
\begin{equation}
  I_{A|B}=  I^{\max}_A+\sum_{ab}p(a,b)\log p(a|b).
\end{equation}

Information and relative information will be the basis of the definition of facts and relative facts. We collect some useful properties of these measures of information in appendix~\ref{sec:info_app}.

\section{Relative Facts}
\label{sec:facts}

\noindent What does it mean for something to be a fact?

In common and in scientific  language, when we know something about the world with sufficient certainty, we say ``it is a fact.'' Accordingly, let us say that the value of a variable $A$ is \textit{a fact} when  the probability distribution $p(a)$ is entirely concentrated on one value, namely when the information $I_A$ is maximal. 

We now introduce another notion of fact based on probability theory: a \textit{relative fact}. We say that the value of $A$ is \textit{a fact relative to} $B=b$ when the conditional probability is concentrated on a single value of $A$, that is if the conditional information $A$ is maximal
\begin{equation}
  I_{A|b}=I^{\max}_A.
\end{equation}
If $A$ is a fact relative to $B$ taking on the value $b$, for all $b$ such that $p(b)\neq0$, then we may simply say that $A$ is a \textit{fact relative to} $B$. In other words, $A$ is a fact relative to $B$ if the relative  information is maximal,
\begin{equation}
\label{rfact}
  I_{A|B}= I_{A}^{\max}.
\end{equation}
Thus we may speak of facts relative to a variable or relative to a variable taking on a specific value. 

Notice that, according to these definitions, the value of a variable $A$ may not be a fact and yet it may be a fact relative to another variable $B$. This happens when $B$ holds enough information about $A$, that is, when ${I_{A}+I_{A:B} = I^{\max}_A}$.

For instance, suppose that you do not know whether Robert is at home ($A=h$) or in his office ($A=o$) and you assign probability 1/2 to the two cases, but you know that Mary knows where Robert is. Then, calling ${B=h}$ and ${B=o}$ the two states of Mary's knowledge, you have ${p(h,h)=p(o,o)=1/2}$ and zero probability for the other two cases. Then \eqref{rfact} holds, we can say that neither Robert being home nor Robert being in his office is a fact, but Robert's location is a fact relative to Mary.  Notice that in this case we know {\em that} Mary knows where Robert is, but we do not know {\em what} she knows about it. We emphasise that nothing here hangs on Mary and Robert being self-reflecting humans. One might make the example with $A$ being Robert's location and $B$ being his smartphone's (given that Robert never leaves his house without it).

Indeed, in common language, when we expect  that our own information about $A$ can increase if we look at $B$, we say that  "$B$ has information about $A$". For instance, we say that an airport panel has information about the arrival of the flights, a geological layer has information about ancient climate, or Mary has information about Robert's whereabouts: in all these cases, we mean that we can get information about a system by looking at another one.   Accordingly, let us say that \textit{the variable $B$ has information about the variable $A$}  if the mutual information $I_{A:B}$ is non-vanishing. In this sense $I_{A:B}$ quantifies how much $B$ "knows" in this general and non-mental sense about $A$: we expect our information about $A$ to increase by accessing $B$.

\smallskip

We now come to a crucial remark. If we have maximal information about a variable $A$, another person could learn its value by asking us about it. In that case, after interacting with us, that person’s information about $A$ will also be maximal. There is no disembodied intelligence or supernatural knowledge: information is physical, it is stored in physical variables and their correlations and can therefore be accessed by other systems. So, whenever we talk about information about some variable or system, it should be understood that we are invariably, although implicitly, talking about information about that variable relative to some other system---generally \textit{us}. According to our definitions, then, a fact is always a fact relative to something, and
\begin{quote}
\textit{Every fact is a relative fact.}
\end{quote}

One further observation is needed before closing this section. Since information and relative information are {\em continuous} quantities, in most realistic cases they will not achieve their maximal value and there will always be some residual uncertainty about the value of a variable. While are many aspects of the world that we estimate we know with near certainty, there is simply no aspect of the world of which we can be {\em totally} certain. Therefore, if we reserve talk about facts to situations when information is truly maximal, we may almost never talk about facts. We may instead use these notions of information as a measure of the definiteness of a fact, and call something a fact when the information is high enough for the level of accessible resolution or confidence needed.

\section{Relational Quantum Mechanics}
\label{sec:RQM}

\noindent The definitions above can be applied in any context in which there is uncertainty about the values of variables measured by probability distributions. In particular, they can be applied directly to quantum theory.

If $A$ is a variable%
\footnote{By \textit{variable} here we mostly mean a quantity represented in the theory by a \textit{self-adjoint operator}, such as the spin of an electron or the values of the different bases of a qubit. The measurement of continuous variables $x$ and $p$ cannot take sharp values, but we can replace their measurements with those of the characteristic functions $\chi_{[x_0,x_1]}$. More generally, one could think of a variable as being represented by a projector-valued measure, as well as repeatable positive-operator-valued measures~\cite[chapter II]{busch1995operational}. The technical details of using POVMs as variables for RQM are developed in~\cite{fano2025relational}.} %
 of a quantum system and $\rho$ is the quantum state of the system, the probability of obtaining the value $a$ if we measure the variable $A$ is given, according to quantum theory, by 
\begin{equation}\label{pd}
  p_\rho(a) = \tr\Pi_a\rho,
\end{equation}
where $\Pi_a$ is the projector on the eigenspace of $A$ corresponding to the eigenvalue $a$.  If $\rho$ is an eigenstate of $A$ with eigenvalue $a$, we can say that $A=a$ is a fact (relative to us), simply meaning that there is no uncertainty on the result of a measurement of $A$, as specified above.\footnote{Note the similarity with the eigenvalue-eigenvector link, often discussed in the philosophy literature, which would state that the system \textit{possesses} the property $A=a$ in this case. Again, we are not making ontological claims. In appendix~\ref{sec:epr}, we compare the definition of relative facts with the EPR criterion for reality~\cite{einstein1935can}.}

Assume now that $A$ and $B$ are two commuting variables of a quantum system. These could belong to two different subsystems or to the same quantum system. When the state of the system is $\rho$, quantum theory then gives the probability distribution for a joint or sequential measurement of $A$ and $B$ as
\begin{equation}
  p_\rho(a,b) = \tr\Pi_a\Pi_b\rho.
\end{equation}
From this distribution $p_\rho$, we can compute the various information quantities defined above, such as $I_A$, $I_B$, $I_{A:B}$, and $I_{A|B}$; we may then apply the definitions of facts and relative facts. If we have maximal information about a variable $A$, namely, if the probability for us to measure $a$ is ${p(a)=1}$, we say that $A=a$ is a fact (for us). 
If the relative information $I_{A|B}$ is maximal, we say that the value of $A$ is a fact relative to $B$, whether or not $A$ is a fact relative to us.

In classical mechanics and classical information theory, all variables commute and therefore all variables could, in principle, simultaneously be a fact. In quantum theory, the variables of a \textit{system} are represented by a non-commutative algebra of operators on a Hilbert space, and only mutually commuting variables may have a joint probability distribution, so only these can simultaneously be a fact according to our definitions. For the rest of the work, we will call a collection of commuting observables a \textit{classical subsystem} of a quantum system. Given a state on an algebra of observables and a \textit{commutative subalgebra} $\mathcal A$  of a quantum system (a classical subsystem), the collection of facts relative to $\mathcal A$ is the \textit{perspective} associated to $\mathcal A$.

\smallskip

Relational quantum mechanics~\cite{rovelli2022relational} is a way of thinking about quantum theory as a theory regarding the information that physical variables have about one another, including the information that \textit{we} have about the world. According to RQM, one may use the apparatus of quantum theory to compute for facts relative to a classical system given other facts relative to that system~\cite{long-thesis}.  This view holds no system as fundamentally different in this regard. All variables and systems are equivalent, and a variable does not have to be special to hold information about another variable. Every quantum state is a quantum state of a system \textit{relative to} some classical (in the sense above) system. As a mathematical object, the quantum state does not encode merely the \textit{beliefs} an agent might have about the world, but the physical relations between the various systems.

The recognition of the role of commuting subalgebras as the carriers of information is a novelty with respect to the RQM literature. In  previous literature (for instance \cite{rovelli2021relational,dibiagio2021stable,dibiagio2022relational,adlam2023information}), there was an attempt to define facts relative to entire quantum systems. This rendered the notion somewhat problematic~\cite{pienaar2021quintet,brukner2021qubits,barbado2025relational}.  Of course, we can always say that a quantum system $\mathcal S$ has information about another quantum system $\mathcal S'$ if one {\em or more} of its commutative subalgebras have information about some variables in $\mathcal S$, and this can be quantified using the \textit{quantum} mutual information. But, defining facts relative to quantum systems leads either to ambiguities as to which variables are a fact relative to a system or to complementary observables being facts simultaneously~\cite{adlam2023information,pienaar2021quintet,brukner2021qubits}. The definition of relative facts based on relative information makes it clear that values of variables can only be coherently thought of as facts relative to {\em commuting} algebras.

This view provides an internally coherent way of thinking about physical phenomena, including quantum phenomena, which is consistent with the language we commonly use to describe them in realistic scientific practice. It does not postulate a physical mechanism for collapse, add equations to standard quantum mechanics, require agents or observers as a primitive notion, nor does it assume a pre-existing classical world. It is close in spirit to Everett's observation that all states we use in quantum physics are relative states \cite{long-thesis}, but it does not give ontological weight to the states themselves. The price to pay (because all interpretations of quantum theory come at a cost for our classical intuitions), is that there are no absolute facts: most statements that are true or false are true or false \textit{in a perspective} and there are no perspective-independent facts; we come back to this in section~\ref{sec:wittgenstein}.

Quantum theory gives us ways of computing joint probability distributions for commuting observables and how these distributions change in time. From these, we can establish what is a fact relative to what. In section~\ref{sec:perspective} we will see how perspectives can agree, disagree, differ,  and merge. But first, we will comment on the role of the observer in RQM.

\section{systems, perspectives, observers}

In quantum theory, we often think of a system as being associated with an entire tensor factor in Hilbert space and all bounded self-adjoint operators as being its observables, especially in finite dimensions. However, a physical system is more appropriately thought of as consisting of a set of physical variables, represented in the theory as distinguished operators on the Hilbert space.
 A spin-$\tfrac12$ is the system represented by (the unital algebra generated by) the $X$, $Y$, and $Z$ observables; a quantum particle by $x$ and $p$, with $[x,p]=i$ and their spectral projectors; and a classical bit is represented by the subalgebra spanned by $\mathds 1$ and $Z$, a subsystem of a qubit.

As Brukner argued, qubits are not observers~\cite{brukner2021qubits}. There are two reasons for this. First, we cannot associate a perspective to a qubit because it has non-commuting variables. Second, while we can associate a \textit{perspective} to a specific variable of the qubit, this is also not \textit{an observer}, because it lacks the resources and structure needed to be called an observer.

Indeed, "perspective" is a wider notion than "observer," in the sense that the latter is a special kind of system we can associate a perspective to. We may associate a perspective to any classical system, any set of commuting observables. Observers, in the standard sense, have additional special properties, variously characterized in different accounts. These often include things like being very heavy so that position and velocity effectively commute (making them classical in the sense of this paper), the ability to store a lot of information, the presence of sufficient decoherence to allow for stable records of the information gathered, and the ability to communicate such information. In some interpretations they are additionally required to be agents capable of taking action and to think about their consequences, like in QBism~\cite{fuchs2019qbism} or Healey's pragmatist reading~\cite{healey2010quantum}. From the point of view of RQM, any Copenhagen or QBism observer is associated with a perspective but not vice-versa, as a perspective could be associated with the spin-$z$ of an electron, which is clearly not an observer.

To be clear, there is nothing wrong in focusing on systems with such properties. Indeed, agents and observers play an essential role in using and understanding quantum theory. What we are pointing out is that there is also a way of understanding quantum theory as a coherent and complete account of natural phenomena that does not require assuming rational agents or a classical world. Interpretations that take these as primitive, then, are simply restrictions to particular cases of specific interest to us humans.

\section{Perspectives differ, but talk}
\label{sec:perspective}

\noindent It is one of the basic consequences of the quantum formalism that different perspectives can disagree on what is or is not a fact \cite{rovelli1996relational}. Yet, perspectives can communicate, and merge. Let us illustrate how this happens with a simple example.\footnote{The analogy with Wigner's friend scenario, treated more carefully in appendix~\ref{sec:WF}, should be evident.}

	Two classical systems $\A$ and $\B$ can come to share their perspectives by interacting properly either with each other or with the systems they are entangled with. Say that $\A$ and $\B$ respectively consist of the single variables $A$ and $B$ of two distinct quantum systems and that they both can interact with a qubit $\S$. Say also that, relative to us, the state of the joint quantum system
\begin{equation}
  \ket{\psi_0}=\ket{+}\ket{a_0}\ket{b_0},
\end{equation}
where $\ket{a_i}$ and $\ket{b_i}$ are eigenstates of $A$ and $B$, respectively, and $\ket{+}=\frac1{\sqrt2}(\ket{0}+\ket{1})$, where $\ket{0}$ and $\ket{1}$ are eigenstates of the variable $Z$ of $\S$.  This implies that two variables $A$ and $B$ are a fact relative to us, and that these two have no information about $Z$. In fact, the value of $Z$ is not a fact relative to us, nor relative to $\A$ or $\B$. Now the systems interact in such a way that $A$ gets correlated with $Z$, resulting in the state
\begin{equation}
 \ket{\psi_1}= \left(\frac1{\sqrt2}\ket0\ket{a_0}+\frac1{\sqrt2}\ket1\ket{a_1}\right)\ket{b_0}.
\end{equation}
The value of $Z$ of is now a fact relative to $\A$ but not relative to $\B$: their perspectives differ.  But $\B$ can come to share $\A$'s perspective by interacting properly with $A$ or $Z$ leading to the state,
\begin{equation}
  \ket{\psi_2}= \frac1{\sqrt2}\ket0\ket{a_0}\ket{b_0}+\frac1{\sqrt2}\ket1\ket{a_1}\ket{b_1}.
\end{equation}
Now $Z$ is a fact relative to both $\A$ and $\B$, as ${I_{Z|A}=I_{Z|B}=I_{Z}^{\max}}.$  Additionally, the value of $A$ is now a fact for $B$, given that ${I_{A|B}=I_A^{\max}.}$ 

The perspectives of $\cal A$ and $\cal B$ have \textit{merged}, in the precise sense that $A$ and $B$ agree on the value of $Z$ and about each other's value, the state of $\S$ conditional on $A=a_i$ is the same as the state of $\S$ conditional on $B=b_i$. This has simply happened via the interaction that has entangled the two systems. 

What would have happened if $\B$ had instead gotten correlated with the $X$ variable of $\S$? The final state of the three systems would have been
\begin{equation}\label{complementary_perspectives}
\ket{\psi_2'}=\frac12\ket{+}\big(\ket{a_0}\!+\!\ket{a_1}\big)\ket{b_0} + \frac12\ket{-}\big(\ket{a_0}\!-\!\ket{a_1}\big)\ket{b_1}.
\end{equation}
In this second scenario, $X$ is a fact relative to $\B$ but not to $\A$. This is to be expected since $\B$ got information about $X$ and $\A$ did not. Second, $A$ is still not a fact relative to $\B$. This happened because $\B$ interacted with a variable that had no information about $A$. Third, $Z$ is not a fact relative to $\A$ anymore, even though $\A$ did nothing in this last timestep. This is because the interaction between $\B$ and $\S$ disturbed $Z$ and changed its correlations with $\A$.

This example illustrates some simple lessons about perspectives and their merging. If the value of a variable $Z$ is a fact relative to another system, and you want to know its value, then interact with either that variable or that system in an appropriate way. If you interact with a complementary variable $X$, then not only your perspectives will not merge, but you might remove a relative fact from other perspectives.

The merging of perspectives offers a solution to RQM's \textit{combination problem}, the question of how to combine the perspectives of various subsystems into the perspective of the super-system~\cite{adlam2024combination,riedel2025composite}. If the variables couple correctly so as to get correlated, their perspectives will agree. This is also the basis for intersubjectivity, as discussed in section~\ref{sec:intersubjective}.

The fact that interactions may remove relative facts reveals what goes wrong in Frauchiger-Renner-type experiments~\cite{frauchiger2018quantum,hausmann2025firewall,aaronson2018its}. When the superobservers start making their measurements on the friends, they change the correlation structure of the variables of the friends. So the kind of inferences the latter make before the supermeasurements are not valid after the supermeasurements.

\section{Measurements are continuous (Intermezzo)}
\label{sec:continuous}

\noindent Quantum "measurements" are sometimes presented as implying a troubling discontinuity.  For instance, if we read the Schr\"odinger's wave function $\psi(\vec x)$ of a non-relativistic particle as a component of a fundamental ontology, its "collapse" at a measurement appears as a mysterious discontinuous breaking of its continuous unitary time evolution: a radioactive atom emits a spherically symmetric  $\psi(\vec x)$, which suddenly collapses onto a much sharper wavefunction centred at the Geiger counter that detected the emitted particle.

A related troubling aspect of textbook quantum mechanics is the question of {\em when} a quantum measurement actually happens. The quantum formalism does not appear, at first sight, to provide a clear answer to this question. 

The tools presented in the previous sections provide an answer and dissolve the confusion.  The evolution of the {\em  information} that a system has about another system always changes {\em continuously}.  During a quantum measurement, the information that a system has about another system is simply  growing continuously. A discontinuity only comes from the definition of "fact" as something whose probability reaches a sufficiently high threshold.

Let us make this precise.  In the conventional von Neumann account of a quantum measurement the measured system and the measuring apparatus are both treated as quantum systems.  Let $A$ be the measured variable, with eigenvalues $a_n$, and $B$ the apparatus' pointer variable, with corresponding eigenvalues $b_k$. Let $b_\mathrm{r}$ be the eigenvalue of the pointer variable corresponding to the initial `ready' state of the apparatus. During the measurement the coupled system evolves from an initial tensor state
\begin{equation}
  \ket{\psi_\mathrm{i}}=\Big(\sum_n \alpha_n\ket{a_n}\Big)\ket{b_\mathrm{r}},
  \label{i}
\end{equation}
to a final entangled state 
\begin{equation}
	\ket{\psi_\mathrm{f}}=\sum_n \alpha_n\ket{a_n}\ket{b_n}.
  \label{f}
\end{equation}
At the beginning of the measurement the mutual information $I_{A:B}$ is zero, and there is no correlation between the variable to be measured and the apparatus.  At the end of the measurement, the relative information $I_{A|B}$ is maximal:  in an ideal measurement the pointer variable gets perfectly correlated to the measured variable.

\begin{figure}[h]
    \centering
    \includegraphics[width=0.9\columnwidth]{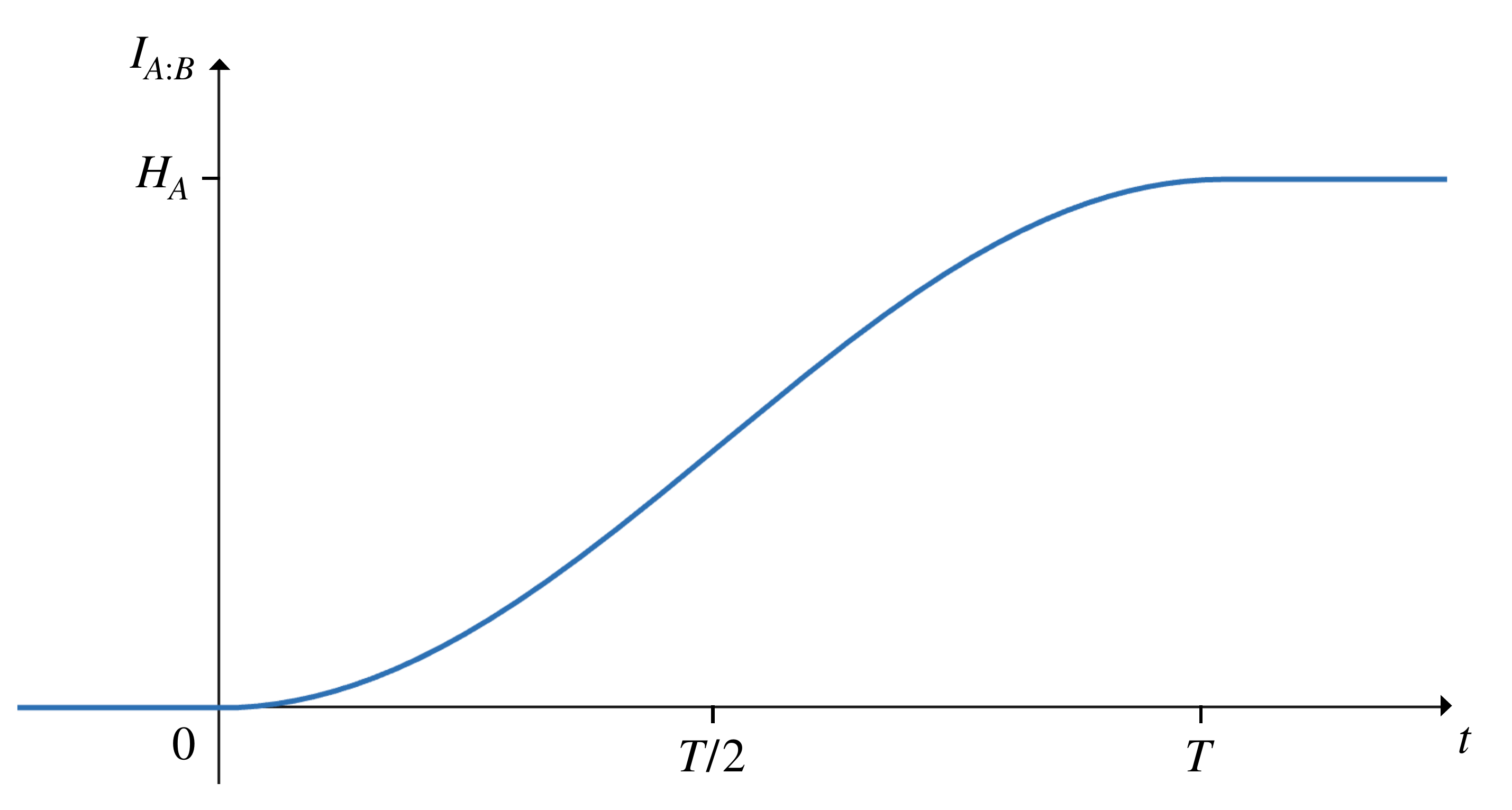}
    \caption{The increase of the apparatus' information about a system in course of a quantum measurement. Before the start of the measurement, the mutual information $I_{A:B}$ vanishes, it then increases continuously to its maximum value $H_A$ during the duration $T$ of the measurement, at which point the relative information $I_{A|B} = I_A+I_{A:B}$ reaches its maximum value.}
    \label{fig:measurement}
\end{figure}

The evolution from the state \eqref{i} to the state \eqref{f} follows  the Schr\"odinger equation and is therefore {\em continuous}.   It follows that the information about the measured variable relative to the apparatus grows {\em continuously}. The apparatus gains information about the measured variable {\em continuously in time, without jumps}, as in figure~\ref{fig:measurement}. There is no special moment in which the information that the apparatus has about the system jumps from  zero to a finite value.  We refer the reader to appendix~\ref{sec:continuous_app} for a more quantitative treatment of such a measurement scheme. 

What does it mean that halfway through a measurement, the pointer variable $B$ has \textit{partial} information about $A$? It means that if we look at the pointer variable, we can infer something incomplete about the result of a subsequent measurement of $A$.  The apparent discontinuity only appears by the requirement of a fact to correspond to achieving \textit{maximum} (or nearly maximum) information.   Since exact certitude is of course impossible in practice, facts always come with some degree of uncertainty.   It is the reduction of this uncertainty to an "acceptable" level that characterizes the end of a von~Neumann measurement and permits us to say that "a variable has been measured" in the conventional sense of quantum measurements in the lab.  

\smallskip

All the above has actually little to do with quantum theory. In classical information theory, measurements of a discrete variable exhibit the same apparent tension between discrete outcomes and continuous information acquisition: while the measurement result needs to be one of a finite set of values, the relative information between the measured variable and the register grows continuously in time~\cite{mackay2003information,cover2006elements}.

\section{Classical world and intersubjective agreement}
\label{sec:intersubjective}

\noindent From the naturalistic standpoint adopted here, knowledge and science are themselves physical phenomena.  When we say that  "we," whether an individual observer or a community sharing information, observe something or know something about the world, we refer to the (commutative algebra of) variables that encode our memories, records, current observations, and our capacity to process and communicate information. The world we describe and know is the perspective associated with this algebra: the collection of facts relative to the variables that constitutes ``we.''  

The classicality and definiteness of our everyday experience follows from the commutative nature of the variables that embody our perspective and the things we are correlated with, and the way these are coupled to each other so as to spread information about themselves in the environment. The way interactions lead to intersubjective agreement has been amply discussed in the work of Zurek and collaborators in their thorough analysis of decoherence, einselection, and quantum Darwinism~~\cite{zurek2003decoherence, zurek2025decoherence}.  Interaction with the decohering environment selects pointer variables---those that are not disturbed while information about  them spreads in the environment. The value of these pointer variables becomes a fact relative to a large plurality of variables of the environment, variables with which many different observers can interact in such a way that the pointer becomes a fact relative to \textit{each of them}. These different observers accessing different fragments of the environment will agree on the value of these variables.

This is largely sufficient for grounding science and for addressing  worries \cite{adlam2022does}  that quantum perspectivism could undermine our possibility of doing science \cite{rovelli2024alicesciencefriendsrelational}.

\section{But what \textit{really} happens?}
\label{sec:wittgenstein}

\noindent Any observation can be accounted for in physical terms as described above. Indeed, all of the above is the physical world. Notice that all this is simple and consistent with common language and laboratory practice.  Anything else beyond the information that physical systems have about one another is outside the domain of empirically accessible physics. When we say that a variable $B$ has information about another variable $A$, we are saying that this is the case \textit{relative to us}. In classical theory, where all variables commute and can take on definite values simultaneously, one can forget this relativity, but this is not the case in our world described by quantum theory. This can feel unsettling, but is it a real problem?

 Every statement is a statement within a perspective and what a system knows about another system is itself a function of what yet another system knows about them, relativity iterates \cite{riedel2024relational}, and one might worry that this leads to an infinite regress. However, the infinite regress is only there if one wishes to exit relationality by demanding a truth beyond any context or perspective. However, exiting relationality is not physics and a truth independent of any relation has no empirical content, as any question is posed and answered by someone or something, and every someone or something is a physical system with a perspective on the world.

In Wigner's friend thought experiment~\cite{wigner1962remarks}, when the friend carries their experiment and measures the value of the qubit, Wigner can assert that the value of the qubit is a fact relative to the friend. Wigner can then go in the lab and interact with the qubit and discover that the value he finds is the same as the one that friend reports having found (see also appendices~\ref{sec:WF}-\ref{sec:ewfs}). One has to resist the temptation to ask ``But what did the friend \textit{really} see?'' That word, ``really,'' stands for ``beyond the information that Wigner, or any other system, could possibly gather about it.'' It  brings the question outside the empirical sphere making it something that cannot be answered by anything or anyone, not even in principle.

Like the questions "How do I know if something \textit{really} moves itself, not relative to something else?"  and "How do I know if two events are \textit{really} simultaneous, not relative to a specific synchronisation convention?", which have been shown to be meaningless by Galileo and Einstein, also the question ``What value did the qubit take, \textit{really}, not relative to friend having had a certain experience?'' is a reflex due to metaphysical beliefs that we can drop.

These questions cannot have an empirical answer, because they ask for information about the world beyond what might be gathered by a physical system.  Questions without empirical content may not have answers, or their possible answers can be arbitrary and not useful. The same way one can add a ``true surface of simultaneity'' to general relativity to provide a global (but unempirical) notion of \textit{now}, we might "complete the ontology" of quantum theory by adding hidden variables such as Bohmian trajectories \cite{sep-qm-bohm} and ``the collection of all relative facts'' \cite{adlam2023information}. These additions are unobservable and are there only to assuage our metaphysical unease; they display weird behaviours such as nonlocality or superdeterminism, as revealed in no-go theorems on the extended Wigner's friend scenario~\cite{brukner2015quantum,brukner2018nogo,bong2020strong,cavalcanti2023consistency,ying2024relating,schmid2023review}; or non-communicating island universes~\cite{pienaar2021quintet} or convivial solipsism~\cite{Zwirn_2016}; or surrealistic Bohmian trajectories \cite{englert1992surrealistic,mahler2016experimental,sole2017surrealistic,hiley2018quantum}.

If we drop these questions, "the riddle does not exist," as Wittgenstein puts it~\cite{wittgenstein2010tractatus,covoni2025tractatus}. We can take quantum phenomena into account, in a fully naturalistic context where no special feature of humans plays a role and where nothing is added to orthodox quantum theory---besides the recognition that any commutative algebra of variables has a perspective associated to it and is a context for communicable facts.

\section{Conclusion}

\noindent We have presented a reformulation of relational quantum mechanics that renders its central ideas quantitatively precise by introducing definitions compatible with standard probability theory and the orthodox quantum formalism. By associating perspectives with commutative subalgebras of observables rather than entire quantum systems, and by defining relative facts in terms of relative information, we have shown how the relational interpretation can be read directly out of the mathematical structure of quantum theory. This reformulation thus responds to criticisms addressed to RQM of relying on vague or not sufficiently quantitative claims without precise mathematical underpinning~\cite{martin-dussaud2023factnets,ormrod2024quantuminfluenceseventrelativity,faglia2025relational,barbado2025relational}. Moreover it allows talk of relative facts also in-between interactions; see also~\cite{fano2025relational}.

Relational quantum mechanics requires no modification to quantum theory, no additional postulates about collapse or consciousness, and no departure from naturalism. We are physical systems among other physical systems, distinguished only by the specific commutative algebra of variables that embodies us and our knowledge. The information we have about the world is a special case of the information that parts of nature have about one another. 

Quantum theory does not describe a view of the world from nowhere, it describes the views from \textit{any}where. Facts are relative to perspectives. Perspectives might disagree but those that couple properly can merge, agree, and share information. The apparently solid, objective world of our experience emerges from the robust agreement between perspectives, selected and maintained by the way the variables interact.

This way of thinking about quantum theory does not fall into mere instrumentalism, as it allows us to reason about what happens beyond our laboratories and direct observations. However, it forsakes a straightforward realism, as it requires us to abandon questions that have no empirical content---questions about what "really" happens beyond the information available to any physical system. In exchange, it offers a coherent account of quantum phenomena that is both mathematically precise and conceptually clear without unnecessary ontological baggage.

\begin{acknowledgements}
\noindent We thank Emily Adlam, Richard Healey, Wayne Myrvold, and Jacques Pienaar for many discussions in the past years, and Anne-Catherine de la Hamette for helpful feedback on these ideas. A special thanks goes to \v{C}aslav Brukner for his insistence, in writing and in person, on the unsatisfactoriness of associating facts with quantum systems; to Niccol\`o Covoni for discussions about Wittgenstein; and to Eric Cavalcanti for pressing, over the years, that the idea of ``the collection of all relative facts'' is not an idea in line with the perspectival spirit of RQM.

ADB's research was funded within the QuantERA II Programme that has received funding from the European Union’s Horizon 2020 research and innovation programme under Grant Agreement No 101017733, and from the Austrian Science Fund (FWF), projects I-6004 and ESP2889224 as well as Grant No. I 5384. This work was also made possible through the support of the WOST, WithOut SpaceTime project (\href{https://withoutspacetime.org}{https://withoutspacetime.org}), led by the Center for Spacetime and the Quantum (CSTQ), and supported by Grant ID 63683 from the John Templeton Foundation (JTF). The opinions expressed in this work are those of the author(s) and do not necessarily refect the views of the John Templeton Foundation.
\end{acknowledgements}

\appendix


\section*{Appendix}

\section{Some properties of information measures}
\label{sec:info_app}

\noindent For any two variables $A$ and $B$:
\begin{itemize}
  \item $I_{A:B}$ is symmetric, $I_{A|B}$ is not.
  \item $I_{A:B}^{\max}=\min\{\log N_A,\log N_B\}$ and in that case either $I_{A|B}$ or $I_{B|A}$ is  maximal.
  \item $I_{A|B}\geq I_A$, and $I_{A|B}^{\max}=I_A^{\max}$
  \item $I_{A:B}>0$ only when neither $A$ nor $B$ is a fact.
  \item $I_{A|B}=I^{\max}_A$ if and only if $I_{A|b}=I^{\max}_A$ for all values of $b$ such that $p(b)\neq0$.
\item $I_{A|A}=I_A^{\max}.$
\end{itemize}
For any $A$, $B$, and $C$ self-adjoint operators on a quantum system:
\begin{itemize}
  \item $I_{A|b}$ is the information about $A$ in the state relative to $B=b$.
  \item If $[A,B]=[A,C]=0$, then $I_A$ takes the same value regardless of whether we measure just $A$ or if we measure it with $B$ or with $C$. 
  \item The mutual information between two variables $I_{A:B}$ of different systems is never larger than the quantum mutual information $\mathcal I$ between the two systems. For pure entangled states, $I_{A:B}\leq \mathcal I/2.$
\end{itemize}

\section{Continuous information gathering}
\label{sec:continuous_app}

\noindent Let us be more quantitative about the point made in section~\ref{sec:continuous}. System starts in the state
\begin{equation}
  \ket{\psi_\mathrm{i}} =\left(\sum_{n=1}^{N_A} \alpha_n\ket{a_n}\right)\ket{b_\mathrm{r}},
\end{equation}
where $\ket{b_\mathrm{r}}$ is one of $N_B>N_A$ possible eigenstates of $B$, representing the "ready" state of the apparatus. We have uncertainty about $A$ measured by ${H_A=\sum_n|\alpha_n|^2\log|\alpha_n|^2}.$ Since the systems are in a tensor state, the mutual information $I_{A:B}^\mathrm{initial}$ vanishes: the apparatus has no information about the system. At the end of the measurement the state is 
\begin{equation}
  \ket{\psi_\mathrm{f}} =\sum_{n=1}^{N_A} \alpha_n\ket{a_n}\ket{b_n}.
\end{equation}
A straightforward calculation shows that the mutual information has become $I^\text{final}_{A:B}=H_A$ and thus, by \eqref{key}, relative information $I_{A|B}$ is now maximal.  How has this happened?

The details will depend on the specifics of the interaction, but since all such interactions feature a continuous evolution from $\ket{\psi_\mathrm{i}}$ to $\ket{\psi_\mathrm{f}}$, an example will illustrate the key points. A dynamics that realises the measurement is given for instance by the Hamiltonian
\begin{equation}
 \hat H = i\,\omega\sum_{n=1}^{N_A} \ketbra{a_n}{a_n} \otimes\big(\ketbra{b_n}{b_\mathrm{r}}-\ketbra{b_\mathrm{r}}{b_n}\big)
\end{equation}
applied in the time interval $t\in[0,T]$, for $T=\pi/2\omega$. During this time we have, for each $n$,
\begin{equation}
  e^{-i\hat Ht}\ket{a_n}\ket{b_\mathrm{r}}=\cos\omega t \,  \ket{a_n}\ket{b_\mathrm{r}}+\sin\omega t \,   \ket{a_n}\ket{b_n},
\end{equation}
so that the state at intermediate times is
\begin{equation}\label{cont_meas_state}
  \ket{\psi(t)}=\cos\omega t \, \ket{\psi_\mathrm{i}}+\sin\omega t \,  \ket{\psi_\mathrm{f}}.
\end{equation}
This leads to the probability distribution
\begin{equation}
 p_t(a_n,b_k) =  |\alpha_n|^2\left(\cos^2\omega t\, \delta_{k\mathrm{r}} + \sin^2\omega t\, \delta_{nk} \right).
\end{equation}
Note that $p_t(a_n)=|\alpha_n|^2$ and so our information about $A$ does not change in time. However, the information $B$ has about $A$ increases as
\begin{equation}
   I_{A:B}(t)=\sin^2(\omega t)\,H_A,
\end{equation}
and correspondingly the relative information grows correspondingly,
\begin{equation}
   I_{A|B}(t)=I^{\max}_A + \cos^2(\omega t) \,H_A.
\end{equation}
The evolution of $I_{A:B}(t)$ is plotted in figure~\ref{fig:measurement}. The quantum measurement is a continuous process of information gathering. 

We note that according to the state $\ket{\psi(t)}$, $A$ is a fact relative to $B=b_n$ for all values of $t>0$, because $p_{t>0}(a|b_n)=\delta_{aa_n}.$ This discontinuous behaviour is an artefact of studying a too idealised system, where $p_0(b_n)$ is exactly vanishing. In particular, any realistic state is always at least a little mixed and full-rank~\cite{guryanova2020ideal,taranto2023landauer}. The probability $p_0(b_n)$ may be very small but it will be non-vanishing. In that case, $I_{A|b_n}=I^{\max}_A-H_A$ at $t=0$ and smoothly increases to close to maximal over a short time. Something similar happens when $B$ is a coarse-graining of a continuous variable.

\section{EPR}
\label{sec:epr}

\noindent We note the similarity of our definition of facts with the famous Einstein-Podolski-Rosen (EPR) definition of "element of physical reality"   \cite{einstein1935can}:
\begin{quote}
\textit{If, without in any way disturbing a system, we can predict with certainty (i.e. with probability equal to unity) the value of a physical quantity, then there exists an element of physical reality corresponding to this physical quantity.}
\end{quote}
 This was presented as a statement about ontology, used to argue that quantum theory offered an incomplete picture of reality. We instead offer a \textit{definition} of "fact" based on probability, information, and predictive capabilities, a definition that matches common usage, and we argue that this definition allows for quantum theory to give a coherent and complete way of thinking about the world.

Let us see what the notion of relative fact has to say about EPR correlations. Consider two entangled qubits in the maximally entangled state
\begin{equation}
  \ket\psi=\frac1{\sqrt2}\ket{00}+\frac1{\sqrt2}\ket{11}=\frac1{\sqrt2}\ket{{+}{+}}+\frac1{\sqrt2}\ket{{-}{-}},
\end{equation}
where $\ket{0},\ket{1}$ are the eigenstates of the $Z$ observable, while $\ket{{\pm}}=(\ket0\pm\ket1)/\sqrt2$ are the eigenstates of $X$.

EPR argue that, by applying their criterion of reality, one is led to claim that both observables $X_2$ and $Z_2$ of the second qubit are simultaneous elements of reality. This is because, if we know that $X_2$ is going to be measured, we can predict the result of that measurement by measuring $X_1$; same goes for $Z_2$ and $Z_1$.

The approach based on relative facts is quite different. The statement is that, while none of the variables is a fact with respect to us, $Z_2$ is a fact relative to $Z_1$ and $X_2$ is a fact relative to $X_1$, in the precise sense that
\begin{equation}
  I_{X_2|X_1}=I^{\max}_{X_2},~~~~
  I_{Z_2|Z_1}=I^{\max}_{Z_2}.
\end{equation}
By interacting with the first qubit appropriately, we can merge our perspective with that of $X_1$ or $Z_1$, thus making the corresponding variable of the second qubit a fact relative to us. However, it is not possible for both $X_2$ and $Z_2$ to be a fact in the \textit{same perspective}.

Note the difference with the GHZ state
\begin{equation}
  \ket{\psi}_\mathrm{GHZ}=\frac1{\sqrt2}\ket{000}+\frac1{\sqrt2}\ket{111},
\end{equation}
where the $Z$ variable of any qubit is a fact relative to the $Z$ variable of any other qubit, but the $X$ variable is not a fact relative to anything. This property of this kind of multipartite entanglement is what solves the ``preferred basis problem'' and leads to the emergence of intersubjectivity~\cite{zurek2025decoherence}, as delineated in section~\ref{sec:intersubjective}.

\section{Wigner's friend}
\label{sec:WF}

\noindent In this appendix we revisit the Wigner's friend \cite{wigner1962remarks}, giving a detailed analysis of the mutual information involved. We'll denote by $\S$, $\F$, and $\W$ the qubit, friend, and Wigner, respectively. As we mentioned above, Wigner and his friend are not entire quantum systems, but classical subsystems. Let's denote by $\H_\S,\H_\F,$ and $\H_\W$ the Hilbert spaces appropriate to model the systems. We will describe the experiment from the perspectives of $\F$ and $\W$ and a \textit{third} observer $\O$, and follow the flow of information.

At $t_0$, $\F$, $\W$, and $\O$ all agree that $\F$ and $\W$ are ready to perform their experiment and all agree on the protocol:
\begin{itemize}\setlength\itemsep{-.3em}
  \item $t_0$: The qubit is prepared in the `+' eigenstate of the $X$ operator.
  \item $t_1$: Friend measures the qubit in the computational basis.
  \item $t_2$: Wigner checks whether Friend completed its measurement.
  \item $t_3$: Wigner asks Friend the result of the measurement.
\end{itemize}

We know that Friend will expect to find the computational basis measurement to yield $0$ or $1$ at $t_1$ with equal chance. To know what will happen to her at $t_2$ we need to know what Wigner will do to her specifically, so let's model Wigner's protocol. 

Wigner assigns to $\SF$ at $t_0$ a state
\begin{equation}\label{psi0}
  \ket{\psi_0} = \ket{+}_\S\ket{\mbox{ready}}_\F\,,
\end{equation}
where $\ket{\mathrm{ready}}_\F$ is the state in $\H_\F$ that is appropriate to descrive  Wigner's friend and her lab (she is awake, her apparata are suitably arranged and so on). A bit later than time $t_1$, $\SF$ will be, relative to Wigner, in the state
\begin{equation}\label{psi1}
\!\! \ket{\psi_1} = \frac1{\sqrt2}\ket0_\S\ket{\mbox{saw $Z{=}0$}}_\F + \frac1{\sqrt2}\ket1_\S\ket{\mbox{saw $Z{=}1$}}_\F\,.
\end{equation}
Again, the states $\ket{\mbox{saw $Z=z$}}_\F$ are states in $\H_\F$ corresponding to definite semiclassical situations appropriate to describe the goings-on in a lab where a definite outcome has happened. Note that the information about about the qubit does not change from $\ket{\psi_0}$ to $\ket{\psi_1}$, it remains minimal, while the information about $\F$ decreases by $\log2$ in that same time, as the friend gets correlated to $Z$. Indeed, the information about $Z$ relative to $\F$ goes from minimal to maximal in that same time, indicating that the value of $Z$ is now a fact relative to $\F$.

At time $t_2$, Wigner checks whether the measurement is complete. He uses his immense experimental knowledge to implement the observable
\begin{equation}
  M = \ketbra{\psi_1}{\psi_1},
\end{equation}
which yields $1$ when $\SF$ is in the state $\ket{\psi_1}$ and $0$ otherwise. Note that $M$ is neither a variable of $\S$ nor $\F$, but a variable living in the rest of the algebra of $\H_\S\otimes \H_\F$. In particular, $M$ does not commute with $\F$. However, when $\W$ measures $M$ on $\ket{\psi_1}$, he obtains $M=1$ while at the same time not affecting any of the relations between $\F$ and $\S$. \textit{Nothing happens to $\F$ during $\W$'s measurement of $M$.} 

Note also Wigner gains no information about $Z$ or $\F$ by performing this experiment. Wigner knows that his friend knows the value of $Z$, that the value of $Z$ is a fact relative to $\F$, but not what value it has. If Wigner wants to know the value of $Z$, there are only two ways of doing so: either asking his friend, or measuring $Z$ directly. Either way, after measuring $Z$ or $\F$, the value of $Z$ will be a fact relative to $\W$ after this measurement.

How does all this look from the perspective of an outside observer? At time $t_0$, $\SFW$ will be in the state
\begin{equation}
  \ket{\Psi_0} = \ket{+}_\S\ket{\mbox{ready}}_\F\ket{\mbox{ready}}_\W
\end{equation}
relative to $\O$. At time $t_1$, this becomes
\begin{equation}
    \ket{\Psi_1} =
    \ket{\psi_1}\ket{\mbox{ready}}_\W.
\end{equation}
Also relative to $\O$, at $t_1$, $Z$ is a fact relative to $\F$. After the $M$ measurement, the state is
\begin{equation}
    \ket{\Psi_2} =
    \ket{\psi_1}\ket{\mbox{saw $M{=}1$}}_\W,
\end{equation}
with $I_Z$, $I_\F$, $I_{Z|\F}$ and $I_{Z\F|\W}$ being the same in $\ket{\Psi_1}$ and $\ket{\Psi_2}.$ When Wigner finally ``opens the box'' at $t_3$ the state becomes
\begin{equation}
	\!\!\begin{aligned}
		\ket{\Psi_3} = 
		&\frac1{\sqrt2}\ket0_\S\ket{\mbox{saw $Z{=}0$}}_\F\ket{\mbox{saw $M{=}1$, $Z{=}0$}}_\W \\
		&+ \frac1{\sqrt2}\ket1_\S\ket{\mbox{saw $Z{=}1$}}_\F\ket{\mbox{saw $M{=}1$, $Z{=}1$}}_\W.
	\end{aligned}
\end{equation}
Relative to $\O$, who has not interacted with $\SFW$ all this time, the uncertainty about $Z$ present at time $0$ has spread to both $F$ and $W$, since now
\begin{equation}
  H_Z=H_\F=H_\W=\log2.
\end{equation}
But note the correlation structure: ${I_{Z|\F}=I_{Z|\W}=I^{\max}_Z}$ and ${I_{\F|\W}=I^{\max}_\F}.$
The value of $Z$ is a fact relative to $\F$ \textit{and} relative to $\W$. What's more, the value of $Z$ relative to $\F$ is the same as the value of $Z$ relative to $\W$. The last statement can be verified by $\O$ by interacting with $Z$, $\F$, and $\W$ in any order. 

In the account of events from both $\W$'s and $\O$'s perspective, $\W$ is able to learn what $\F$ knows by interacting properly with her. From $\W$'s perspective, this follows from the fact that QM predicts that the measurement on $\F$ at $t_3$ will yield perfect information about $Z$. In $\O$'s perspective, it follows from the correlations between $Z,$ $\F,$ and $\W$ in the state $\ket{\Psi_3}$, which again, don't say anything more than, were $\O$ to measure $\W$, the interaction will yield perfect information about $Z$ \textit{and} $\F$. Note in either case, it makes no sense for $\W$ or $\O$ to ask ``but what did $\F$ \textit{really} see?'' beyond the answer they could obtain by interacting with $\F$  directly.

\section{Extended Wigner's Friend Scenario}
\label{sec:ewfs}

Finally, let us make a comment about the status of the assumption of \textit{absoluteness of observed events} (AOE) according to RQM. To do so, let us consider the minimal extended Wigner's friend scenario (EWFS)~\cite{brukner2020facts,bong2020strong,schmid2023review}.

In the minimal EWFS, Alice plays the role of Wigner and has arbitrary quantum control on her friend Charlie. On each round of the experiment, Charlie measures the same variable of a quantum system $\S$. On some rounds, Alice ``opens the box'' and asks Charlie the outcome of his measurement. On other rounds, Alice may perform an interference experiment. The twist over the WF scenario is that $\S$ is entangled with another system $\S'$ that another observer, Bob, measures.

We consider the statistics $f(ab|xy)$ of the outcomes of Alice's and Bob's  measurements given their respective measurement choices. AOE states that since Charlie makes a measurement on each round, the statistics observed by Alice and Bob are the marginal of an underlying probability distribution involving Charlie's outcome,
\begin{equation}\label{aoe1}
  f(ab|xy) = \sum_c p(abc|xy).
\end{equation}
Additionally, AOE also states that when Alice opens the box, she \textit{learns} what Charlie saw, that is
\begin{equation}\label{aoe2}
  p(ac|x{=}1)\propto\delta_{ac},
\end{equation}
where $x=1$ corresponds to the measurement choice of asking Charlie their result. By adding assumptions of no-superdeterminism and locality, one can derive bounds on the statistics $f(ab|xy)$, the so-called local-friendliness inequalities~\cite{bong2020strong}. Quantum theory however allows Alice and Bob to violate these inequalities and, since the no-superdeterminism and locality assumptions are the same as those used in Bell's first theorem~\cite{bell1964einstein}, one is pushed to deny AOE~\cite{brukner2020facts,cavalcanti2021implications,dibiagio2025bell}.

But what does denying AOE actually mean? The perspective of RQM is the following.

When Alice opens the box, she sees what Charlie saw, in the precise sense that Bob, or anyone else, can ask Alice and Charlie what they saw and they will give the same answer, then Bob can check $\S$ itself and any relevant part of Alice's or Charlie's lab, and everything will agree. This is guaranteed by quantum theory and there is no other empirically meaningful sense in which ``Alice sees what Charlie saw'' could be true, as discussed in section~\ref{sec:wittgenstein}. That is, equation~\eqref{aoe2} is true according to quantum theory.

Things are more subtle when Alice chooses to perform an interference experiment. Before Alice does her measurement, the outcome of Charlie's measurement is a fact relative to Charlie. However, when she does her experiment, she forsakes the possibility of ever learning about Charlie's outcome before her measurement. There is no sense of postulating a single value for $C$.

Note also that knowledge of the results of Alice, Bob, and Charlie's experiment may be stored in mutually commuting observables $A$, $B$, and $C$, and quantum theory will allow to compute a probability $p_\mathrm{QM}(abc|xy)$ at all moments of the experiment. However, at no time will $p_\mathrm{QM}$ satisfy~\eqref{aoe1}, that is
\begin{equation}\label{aoe3}
  f(ab|xy) \neq \sum_c p_\mathrm{QM}(abc|xy),
\end{equation}
for $x{\neq}1$. Charlie's outcome is not a \textit{stable fact} with respect to Alice, in the RQM sense of \cite{dibiagio2021stable} precisely because Alice is assumed to have complete quantum control over Charlie. The failure of normal probability theory, the interference terms, is a symptom of the relationality of facts; see also~\cite{soulas2023measurementproblemlighttheory}.

\bibliography{refs.bib}
\end{document}